\pdfoutput=1

\documentclass[11pt,a4paper]{article}
\usepackage{jinstpub}
\usepackage{booktabs}

\newcommand{\FDG}{\ensuremath{^{18}}F}

\title{Investigation of the Coincidence Resolving Time performance of a PET scanner based on liquid xenon: A Monte Carlo study}
\author[a]{J.J.~Gomez-Cadenas,}
\author[a]{J.M.~Benlloch-Rodr\'iguez,}
\author[a,1]{P.~Ferrario,}\note{Corresponding author.}
\emailAdd{paola.ferrario@ific.uv.es}
\author[b]{F.~Monrabal,}
\author[a]{J.~Rodr\'iguez,}
\author[c]{J.F. Toledo}
\affiliation[a]{Instituto de F\'isica Corpuscular (IFIC), CSIC \& Universitat de Val\`encia,\\
Calle Catedr\'atico Jos\'e Beltr\'an, 2, 46980 Paterna, Valencia, Spain}
\affiliation[b]{Department of Physics, University of Texas at Arlington\\
Arlington, Texas 76019, USA}
\affiliation[c]{Instituto de Instrumentaci\'on para Imagen Molecular (I3M), Universitat Polit\`ecnica de Val\`encia\\ 
Camino de Vera, s/n, Edificio 8B, 46022 Valencia, Spain}

\abstract{The measurement of the time of flight of the two 511 keV gammas recorded in coincidence in a PET scanner provides an effective way of reducing the random background and therefore increases the scanner sensitivity, provided that the coincidence resolving time (CRT) of the gammas is sufficiently good. The best commercial PET-TOF system today (based in LYSO crystals and digital SiPMs), is the VEREOS of Philips, boasting a CRT of 316 ~ps (FWHM). 

In this paper we present a Monte Carlo investigation of the CRT performance of a PET scanner exploiting the scintillating properties of liquid xenon. We find that an excellent CRT of 70 ps (depending on the PDE of the sensor) can be obtained if the scanner is instrumented with silicon photomultipliers (SiPMs) sensitive to the ultraviolet light emitted by xenon. Alternatively, a CRT of 160 ps can be obtained instrumenting the scanner with (much cheaper) blue-sensitive SiPMs coated with a suitable wavelength shifter.
These results show the excellent time of flight capabilities of a PET device based in liquid xenon.   
}


\begin{document}

\maketitle

\section{Introduction}
Positron Emission Tomography (PET) is a non invasive imaging technique that produces a three-dimensional image of functional processes --it does not show anatomic features, but it rather  measures the metabolic activity of the cells-- in the body. PET is used in
both clinical and pre-clinical research to study the molecular bases and treatments of
disease. The principle of operation relies in injecting into the patient a  
biologically active molecule doped with a radioactive isotope, called tracer (a standard tracer is fluorodeoxyglucose, formed substituting an atom of oxygen by the isotope \FDG\ in a glucose molecule). The radionuclide decays and the resulting positrons
subsequently annihilate with electrons after traveling a short distance within the subject.
Each annihilation produces two 511 keV photons emitted most of the time in opposite
directions and these photons are registered by a detection system. For a pair coincidence event, if the energy of two photons stays within a
pre-set energy window, centred on the 511 keV photopeak, and the timing difference stays within a pre-set time window, a coincidence event will be registered and constitutes a line-of-response (LOR) for image
reconstruction.

The reconstruction of the image in a PET system requires crossing many LORs which in turn define one emission point in the area under study. A LOR can be formed by: 1) a {\bf true coincidence}, which occurs when both photons from an annihilation event are detected, neither photon undergoes any form of interaction prior to detection, and no other event is detected within the coincidence time-window; 2)
a {\bf scattered coincidence}, which occurs when one or both detected photons undergo at least one Compton scattering interaction prior to detection;
3)  a {\bf random coincidence}, which occurs when two photons, not arising from the same annihilation event, impinge the detectors within the coincidence time window of the system. Scatter and random coincidences add a background to the true coincidence distribution, decreasing contrast and causing the isotope concentrations to be overestimated. They also add statistical noise to the signal. 

The measurement of the time of flight (TOF) of the two 511 keV gammas recorded in coincidence in a PET scanner provides an effective way of reducing the background due to random coincidences and therefore increases the scanner sensitivity, provided that the coincidence resolving time (CRT) of the gammas is sufficiently good. Existing commercial systems based in LYSO crystals, such as the GEMINI \cite{gemini}, and most recently the 
VEREOS \cite{vereos} (both of them manufactured by Philips) reach CRT values of $\sim$ 600 (300) ps (FWHM). 

Recent research, using small setups, has found even better results. For example Ferri et al. \cite{LysoFBK} report measurements using a pair of small LYSO crystals 
($3 \times 3 \times 5 {\rm\ mm^3}$ )  read out by high-fill factor SiPMs. With this setup, a CRT of $157\pm 5$ ps was found at ambient temperature ($20\,^{\circ}{\rm C}$). The CRT was considerably improved working at $-20\,^{\circ}{\rm C}$, to a value of
 $125\pm 5$ ps. The limiting factor of the CRT at ambient temperature was the impact of the dark current rate (DCR). Gundacker and co-workers have found a CRT of $85 \pm 4 $ ps for very small crystals of $2 \times 2 \times 3 {\rm\ mm^3}$  \cite{LysoCRT}.

In this paper we present a Monte Carlo investigation of the CRT performance of a recently proposed new type of
PET scanner, called PETALO (Positron Electron TOF Apparatus using Liquid xenOn). 
 In section \ref{sec.LXe} we summarize the relevant properties of liquid xenon (LXe) as scintillating material, review  the mechanisms which produce scintillation light in LXe and propose a parameterization of the measured scintillation data in terms of two decay constants. In section \ref{sec.PET} we present the main features of the PETALO scanner. In section \ref{sec.CRT}  we discuss the CRT performance in LXe cells using SiPMs sensitive to the xenon ultraviolet  light and conventional SiPMs coated with a suitable wavelength shifter. Conclusions are presented in section \ref{sec.conclu}.  

%

\section{Liquid xenon as detection material}
\label{sec.LXe}

\subsection{Physical properties of liquid xenon relevant for PET}

\begin{figure}[!bhtp]
	\centering
	\includegraphics[scale=0.6]{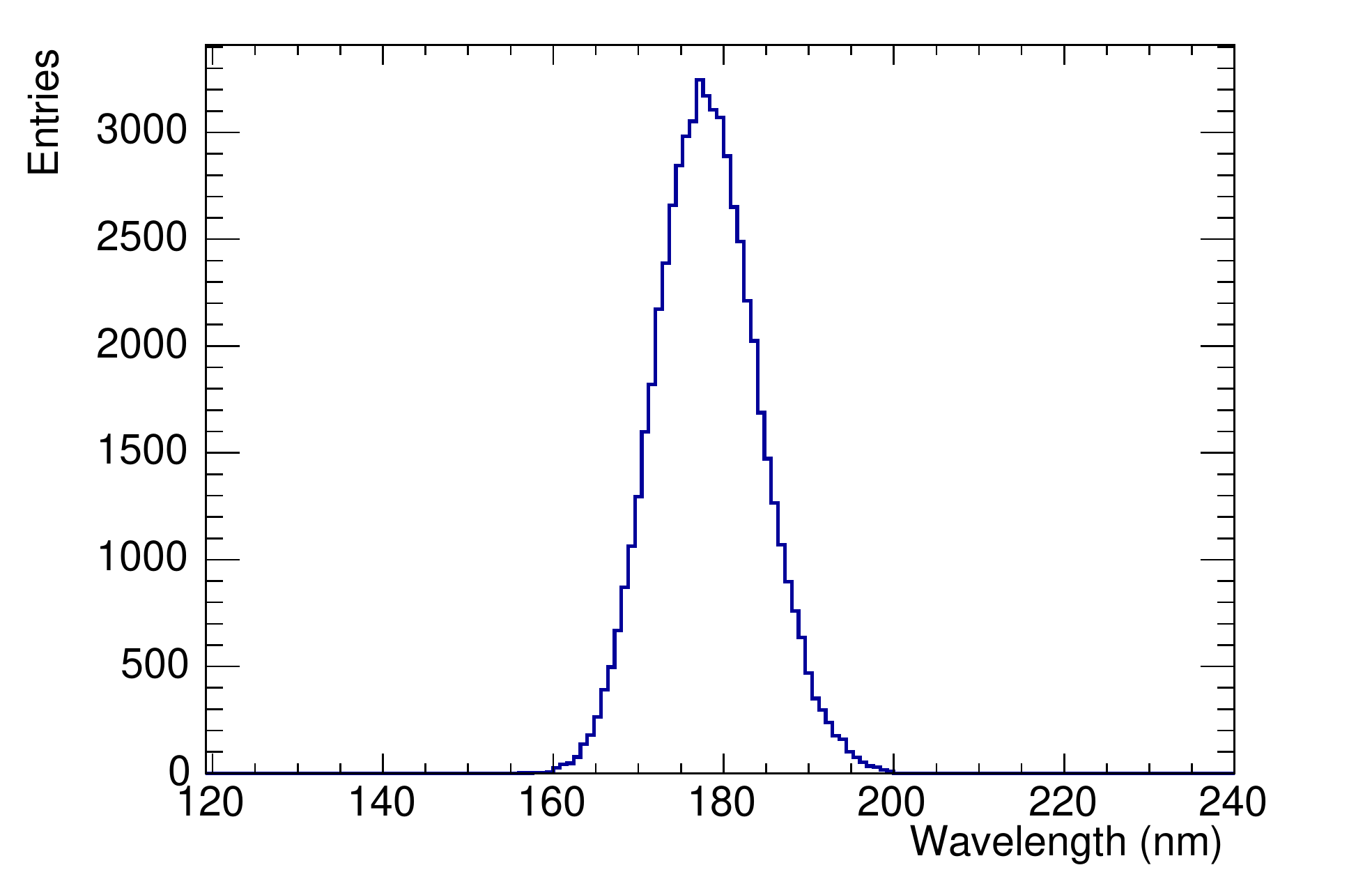}
	\caption{\label{fig.spectrumLXe} Emission spectrum of scintillation photons in LXe, simulated as a gaussian distribution with mean 178 nm and FWHM 14 nm \cite{EmissionLXe}.}
\end{figure}

Xenon is a noble gas. It responds to the interaction of ionizing radiation producing about 60 photons per keV of deposited energy \cite{Chepel:2012sj}. The emitted photons have wavelengths in the ultraviolet range 
(figure \ref{fig.spectrumLXe}) with an average wavelength of 178 nm. The scintillation signal is fast 
(see section \ref{sec.scint}) and can thus result in an excellent CRT.  In its liquid phase (the triple point of xenon is reached at a temperature of 161.35 K and a pressure of 0.816 bar \cite{TempLXe}) xenon has a reasonably high density (2.98 g/cm$^3$)  \cite{DensityLXe}, which gives an attenuation length of 3.7 cm for 511-keV gammas \cite{AttLengthLXe}  and an acceptable Rayleigh scattering length (36.4 cm) \cite{RayleighLXe}, which makes it suitable for PET applications. Its main attractive features are:

\begin{enumerate}
\item {\bf A high scintillation yield}  ($\sim$ 30\,000 photons per 511 keV gamma). 
\item {\bf LXe is a continuous medium with uniform response}. Therefore, the design of a compact system is much simpler than in the case of solid detectors of fixed shape. It is also possible to provide a 3D measurement of the interaction point, and, thus, a high resolution measurement of the depth of interaction (DOI). Furthermore, in LXe it is possible to identify Compton events depositing all their energy in the detector as separate-site interaction, due to its relatively large interaction length.  
Once an event in the region of interest (around 511 keV of total deposited energy) is identified, the pattern of recorded light on SiPMs can in principle be inspected to find one or more depositions, using, for instance, neural networks \cite{tfm}. This increases the sensitivity of the system, since those events can, in principle, be used for image reconstruction. 
\item {\bf The temperature at which xenon can be liquefied at a pressure very close to the atmospheric} is high enough as to be reached using a basic cryostat. Also, at this temperature SiPMs can be operated normally, and their DCR is essentially negligible (see, for instance figure 17, in reference \cite{dcr}). 
\item {\bf The cost} of LXe is reasonable, around 3 \$/cc. 
 \end{enumerate}

\subsection{Scintillation in LXe} \label{sec.scint}

When a 511 keV gamma interacts in liquid xenon it will produce a secondary electron (by either photoelectric effect, which happens in $\sim$ 20 \% of the events, or Compton interaction) which in turn will propagate a short distance in the liquid, ionizing the medium. Most of the scintillation light in xenon is emitted by diatomic excited molecules which are formed through two distinct processes:
\begin{enumerate}
\item Excitation of atoms by electron impact with subsequent formation of strongly bound diatomic molecules in the excited state (excitons). We refer to this process as scintillation due to exciton self-trapping or SE.
\item Recombination of the ionization electrons with positive ions. We call this process scintillation due to recombination or SR. 
\end{enumerate}

Both processes are very fast and therefore scintillation in LXe does not have a physical rise constant. The two decay constants are related with the decay of the two lowest excited molecular states 
(${}^1{\Sigma_u^+},{}^3{\Sigma_u^+}$). The decay of the singlet state 
(${}^1{\Sigma_u^+}$) occurs with a lifetime $\tau_1 =2.2$ ns, while the lifetime of the triplet state (${}^3{\Sigma_u^+}$) is $\tau_2 =27$ ns \cite{Kubota79}.

The relative contribution of SE and SR processes to the LXe scintillation light has been measured to be \cite{Kubota79}:
\begin{equation}
I = 0.3\times I^S + 0.7\times I^R
\label{eq.tot}
\end{equation}
where $I^S$~refers to the intensity of the SE process and $I^R$ to the intensity of the SR process.

Also, after the measurements in reference \cite{Kubota79}, $I^S$ and $I^R$ can be written as:
\begin{eqnarray}
I^S & = & 0.15\times I_1^S(t) +  0.85\times I_2^S(t) \\
I^R & = & 0.44\times I_1^R(t) +  0.56\times I_2^R(t)
\label{eq.1}
\end{eqnarray}

The time dependence of the SE process is described by:
\begin{equation}
I^S_i = \frac{e^{-t/\tau_i}}{\tau_i} 
\label{eq.2}
\end{equation}
where $i=1,2$. The time dependence of the RE process is more complicated, since the recombination time $T_r$
of the electrons is slow compared with the fast constant $\tau_1$. Then, $I^R_1$~can be parametrized as: 

\begin{equation}
I^R_1 =(1 + \frac{t}{T_r})^{-2}
 \label{eq.3}
\end{equation}
where $T_r = 15$ ns for xenon. On the other hand, since $\tau_2 > T_r$, $I^R_2$ can be described as in the case of the SE processes: 

\begin{equation}
I^R_2 =\frac{e^{-t/\tau_2}}{\tau_2}
 \label{eq.4}
\end{equation}
\begin{figure}[!bhtp]
	\centering
	\includegraphics[scale=0.6]{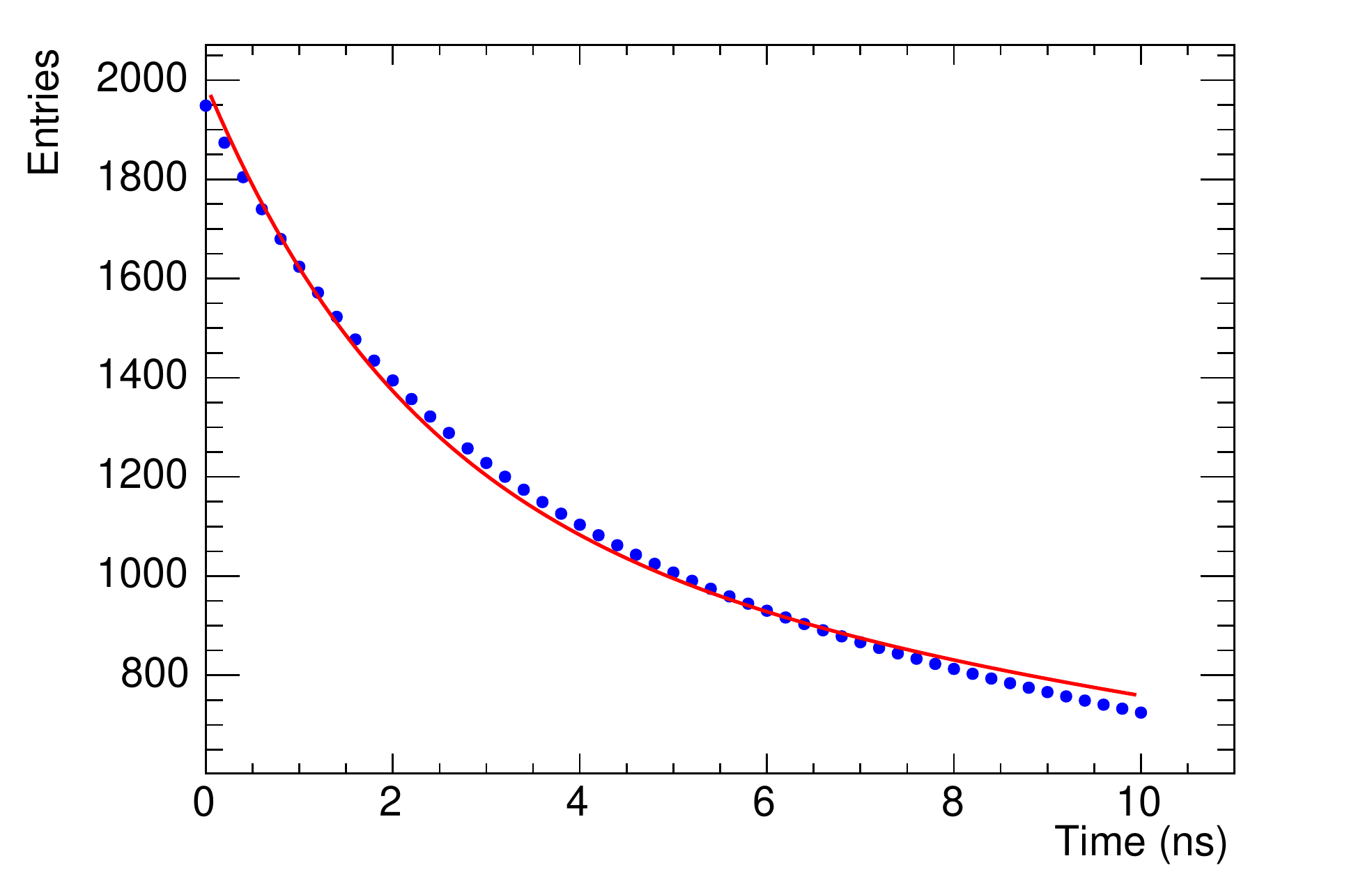}
	\caption{\label{fig.scint} Intensity of scintillation in LXe as a function of time for the first 10 ns following a 511 keV gamma interaction. Blue dots are obtained from equation \eqref{eq.tot.full}, using time constants and intensities from reference \cite{Kubota79}. The red line is the same data fitted with the function in equation \eqref{eq.scint}. Although the distribution extends up to several hundreds of nanoseconds, only the first 10 ns are shown, which is the range used for the fit. In the simulation studies presented in this work, only the first nanoseconds are relevant, since the CRT is determined by the arrival time of first few photoelectrons.}
\end{figure}
Thus, equation \eqref{eq.tot} can be expressed as:
\begin{equation}
I = 0.3\times 0.15\times \frac{e^{-t/\tau_1}}{\tau_1} +  (0.3 \times 0.85 +  0.7\times 0.56) \times \frac{e^{-t/\tau_2}}{\tau_2} + 0.7\times 0.44\times (1 + \frac{t}{T_r})^{-2}
 \label{eq.tot.full}
\end{equation}

Figure \ref{fig.scint} shows the total intensity of scintillation in LXe for a 511 keV gamma (resulting in a total of $N_\gamma =30\,000$ VUV photons), for the first 10 ns after the interaction, according to equation \eqref{eq.tot.full} (blue dots). During this initial period, this distribution is well described in terms of the sum of two exponential distributions with decay constants $\tau_1$ and $\tau_2$. The fit, shown in figure \ref{fig.scint}, gives the following distribution for the number of photons: 

\begin{equation}
I(t) = N_\gamma (0.07 \times \frac{e^{-t/\tau_1}}{\tau_1} + 0.93 \times \frac{e^{-t/\tau_2}}{\tau_2})
\label{eq.scint}
\end{equation}
Equation \eqref{eq.scint} fits better in the short range, which is the most relevant for CRT estimations. This parametrization is the one used in the simulation studies throughout this paper, since the Geant4 simulation toolkit on which our simulation is based \cite{Agostinelli:2002hh} can only model the production of scintillation photons as a single or double exponential.

\section{PETALO}\label{sec.PET}

\subsection{The PETALO concept}


The possibility of building a LXe PET based on the excellent properties of LXe as scintillator was first suggested by Lavoie in 1976 \cite{lavoie}. In 1993, Chepel proposed to read both scintillation and ionization charge in a LXe Time Projection Chamber \cite{chepelFirst} and an extensive work has been done since then by him and co-workers \cite{chepelLXe, chepelRes, chepelEnergyRes}. In the late '90s, Doke and Masuda proposed to use a liquid xenon scintillating calorimeter read out by photomultipliers surrounding completely the whole volume \cite{DokeMasuda}. This idea, applied to PET, was studied by the Waseda group \cite{Doke1,Nishikido2,Nishikido1} with a prototype based in LXe cells read out by VUV-sensitive PMTs which covered 5 of the 6 sides of the cell. 

PETALO (Positron Electron TOF Apparatus using Liquid xenOn) is a proposed new type of PET scanner which exploits the copious scintillation of liquid xenon and the availability of state-of-the-art SiPMs and fast electronics designed to maximize TOF performance \cite{Petalo2015}. The detector building block is the liquid xenon scintillating cell (LXSC). The cell shape and dimensions can be optimised depending on the intended application. In particular, the LXSC2 instruments the entry and exit faces of the box (relative to the gammas line of flight) with silicon photomultipliers, while all the other faces are covered with a reflecting material such as PTFE. They are read out by application-specific integrated circuits (ASICs) optimized for excellent timing resolution. PETALO is a compact, homogenous and highly efficient detector which shares many of the desirable properties of monolithic crystals, with the added advantage of high yield and fast scintillation offered by liquid xenon, low noise due to cryogenic operations which virtually eliminates the SiPMs DCR, and the potential of low cost. 

The energy resolution $R$ of a liquid xenon scintillation detector is a combination of light collection variation due to the detector geometry $R_g$, the statistical fluctuation of number of photoelectrons from the sensors $R_s$, the fluctuation of electron-ion recombination due to escape electrons $R_r$, and the intrinsic resolution from liquid xenon scintillation light $R_i$, due to the non-proportionality of scintillation yield, also associated to fluctuations in the number of secondary electrons. Therefore:

\begin{equation}
R^2 = R_g^2 + R_s^2 + R_r^2 + R_i^2
\end{equation}

The contribution of the intrinsic terms, ($R_r$~and $R_i$) has been measured to be 11 \% FWHM \cite{aprileRes} for 511 keV gammas. The contribution of photoelectron statistics and geometrical effects for the LXSC2 configuration was found to be 3 \% FWHM \cite{tfm}. The combination of both effects yields an expected 
overall resolution of the order of 12\% FWHM for 511 keV gammas, similar to that of current commercial detectors. 

The spatial resolution in the (x,y) coordinates (transverse to the gammas line of flight), meaning the precision in identifying the position of a single-vertex interaction, is obtained in the LXSC2 by the weighted pulses in the SiPMs using a centre-of-gravity algorithm. With this procedure, a resolution of 4.7 mm FHWM has been obtained in Monte Carlo simulations for a box of $5\times5\times5$ $\textrm{cm}^3$ \cite{tfm}. In the current study we have a smaller pitch (there is no space between SiPMs), so a slightly better $xy$ resolution is expected. The DOI is obtained by computing the ratio between the total signal recorded in the entry and exit face, and its resolution is found to be 3 mm FWHM \cite{tfm}. Both resolutions are an average for interactions occurring in the full detector volume, without fiducial cuts.

\subsection{VUV-sensitive SiPMs versus SiPMs coated with TPB}


%
\begin{figure}[!htbp]
	\centering
	\includegraphics[scale=0.5]{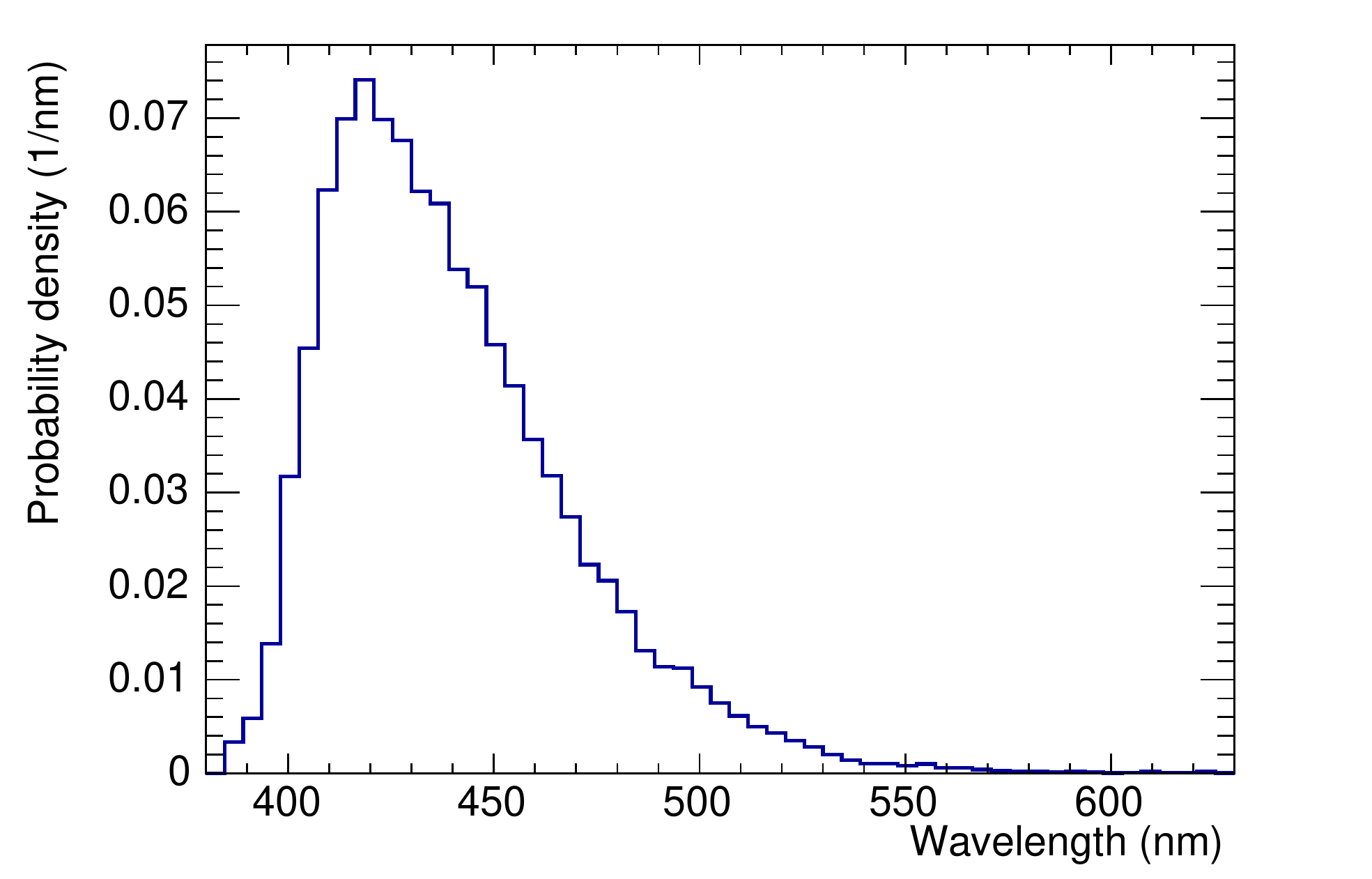}
	\caption{Visible re-emission spectrum of TPB, simulated according to measurements from reference \cite{Gehman:2011xm}.}
\label{fig.tpb} 
\end{figure}

%

The scintillation light of LXe peaks around 178 nm (VUV region). Therefore the LXSC2 needs to be instrumented either with VUV-sensitive SiPMs (as for example the upgraded LXe calorimeter of the 
MEG experiment \cite{Ogawa:2015ucj}) or by conventional SiPMs coated with a wavelength shifter such as tetraphenyl butadiene (TPB) (as for example in the NEXT experiment \cite{Alvarez:2013gxa}). 
The VUV-SiPMs tested for the MEG as well as for the future nEXO 
experiment \cite{Ogawa:2015ucj,Ostrovskiy:2015oja,pdeVUV} reach a PDE between 
15\% and 20\%.  Conventional SiPMs can reach today a PDE of around 50\% \cite{pde}. When they are coated with a thin layer of TPB, the VUV light is absorbed by the wavelength shifter with an efficiency of 80\%,  
shifted to $\sim$ 430 nm 
(figure \ref{fig.tpb}) 
and re-emitted isotropically \cite{Gehman:2011xm}. The decay constant of TPB has been measured  in thin films \cite{TPBtau} and a value of 2.2--3 ns is found, depending on the TPB concentration. %

\section{Monte Carlo study of the CRT in the LXSC2} 
\label{sec.CRT}

\begin{figure}[!bthp]
	\centering
	\includegraphics[scale=0.35]{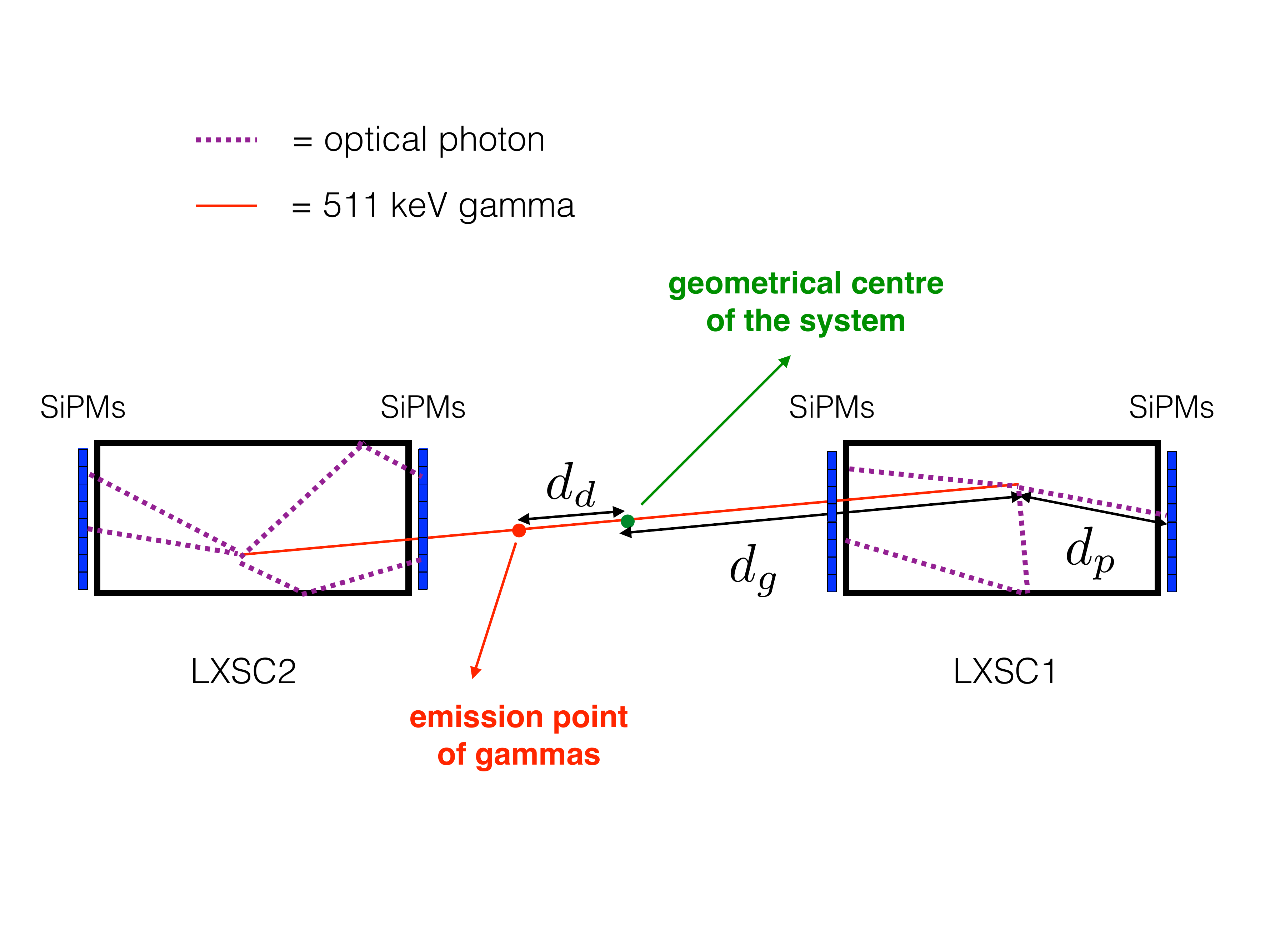}
	\caption{\label{fig.psetup} Scheme of the Monte Carlo simulation set-up, where $d_g$ is the distance between the geometrical centre of the system and the interaction point of the 511-keV gamma, $d_d$ is the distance between the geometrical centre of the system and the emission point of the gammas and $d_p$ is the distance that the scintillation VUV photon covers between its emission and detection point. }
\end{figure}

To systematically study the different factors that enter the CRT, we have simulated a LXe setup and a LYSO setup, using the Geant4 toolkit \cite{Agostinelli:2002hh}, with distribution version number 10.1.p01.  All the simulations described in this paper use this set-up. The LXe setup consists of two LXSC2 of 
$2.4 \times 2.4 \times 5 {\rm ~cm^3}$, facing each other in opposite sides of a 511 keV gamma source and equipped with SiPMs which can be either sensitive to the xenon ultraviolet (VUV) light, or coated with a wavelength shifter such as tetraphenyl butadiene (TPB). The active area of the SiPMs is $3\times 3$ $\textrm{mm}^2$, thus each instrumented face has 64 sensors covering the whole area at a pitch of 3 mm. The uninstrumented faces reflect optical photons according to a lambertian distribution with an efficiency of 97 \%. Rayleigh scattering is simulated, with an interaction length of 36 cm. In Table \ref{thetable} the Monte Carlo set-up specification as well as the values of the relevant parameters used are summarized.

\begin{table*}[h!]\centering
\begin{tabular}{@{}l|l@{}}\toprule
Parameter & Value \\  \midrule
\multicolumn{2}{c}{Geometry}\\  \midrule
Cell dimensions & $2.4\times 2.4 \times 5$ cm$^3$ \\
Distance between cell entry faces & 20 cm \\
SiPM active area &  $3\times 3$ $\textrm{mm}^2$\\
SiPM pitch & 3 mm\\
Number of SiPMs per face & 64 \\
Reflectivity of PTFE walls & 0.97 \\
\midrule
\multicolumn{2}{c}{LXe physics properties}\\  \midrule
Density & 2.98 g/cm$^3$ \\
Attenuation length for 511 keV gammas & 3.7 cm \\
Scintillation yield for 511 keV gamma & $\sim 30\ 000$ photons \\
Average scintillation wavelength & 178 nm \\
Rayleigh scattering length &  36.4 cm \\
 \end{tabular}
\caption{Summary of the Monte Carlo set-up specifications and the LXe relevant properties used in the simulations.}\label{thetable}
\end{table*}


%

  Back to back 511 keV gammas are shot isotropically, from a common vertex within the solid angle covered by two opposite LXSC2 ($2.4 \times 2.4 \times 5\ {\rm cm^3}$). The gammas interact in the LXSCs producing scintillation photons according to equation \eqref{eq.scint}  which are propagated through the material. The photons reach eventually the SiPMs and, depending on the PDE of the SiPM, may result in a photoelectron.
 
We denote $d_g$ as the distance from the centre of the geometrical system to the interaction vertex of each 511 keV gamma, $d_d$ as the displacement of the gamma emission vertex from the centre of the geometrical system and $d_p$ as the distance from the interaction vertex to the detection vertex (i.e., the position of the sensor), as illustrated in figure \ref{fig.psetup}. If $t_1,t_2$ are the time of the first photoelectron recorded in each one of the LXSC, the time difference between them can be written as:
\begin{equation}
t_1 - t_2 = 2\frac{d_d}{c} + \frac{\Delta d_g}{c} + \frac{ \Delta d_p}{v_p}
\label{eq.first}
\end{equation}
where $v_p$ is the velocity of the scintillating photon. Therefore, the difference in time between the gamma emission vertex and the centre of the geometrical system can be expressed as:
\begin{equation}
\Delta t  \equiv \frac{d_d}{c}  =  \frac{1}{2}(t_1 - t_2 - \frac{\Delta d_g}{c} - \frac{\Delta d_p}{v_p}) 
\label{eq.CRT}
\end{equation}
%
%
The CRT,
$\delta \Delta t$, is defined as the variance\footnote{expressed in FWHM unless explicitly stated otherwise.} of the $\Delta t$ distribution. 

\begin{figure}[!bhtp]
	\centering
	\includegraphics[scale=0.6]{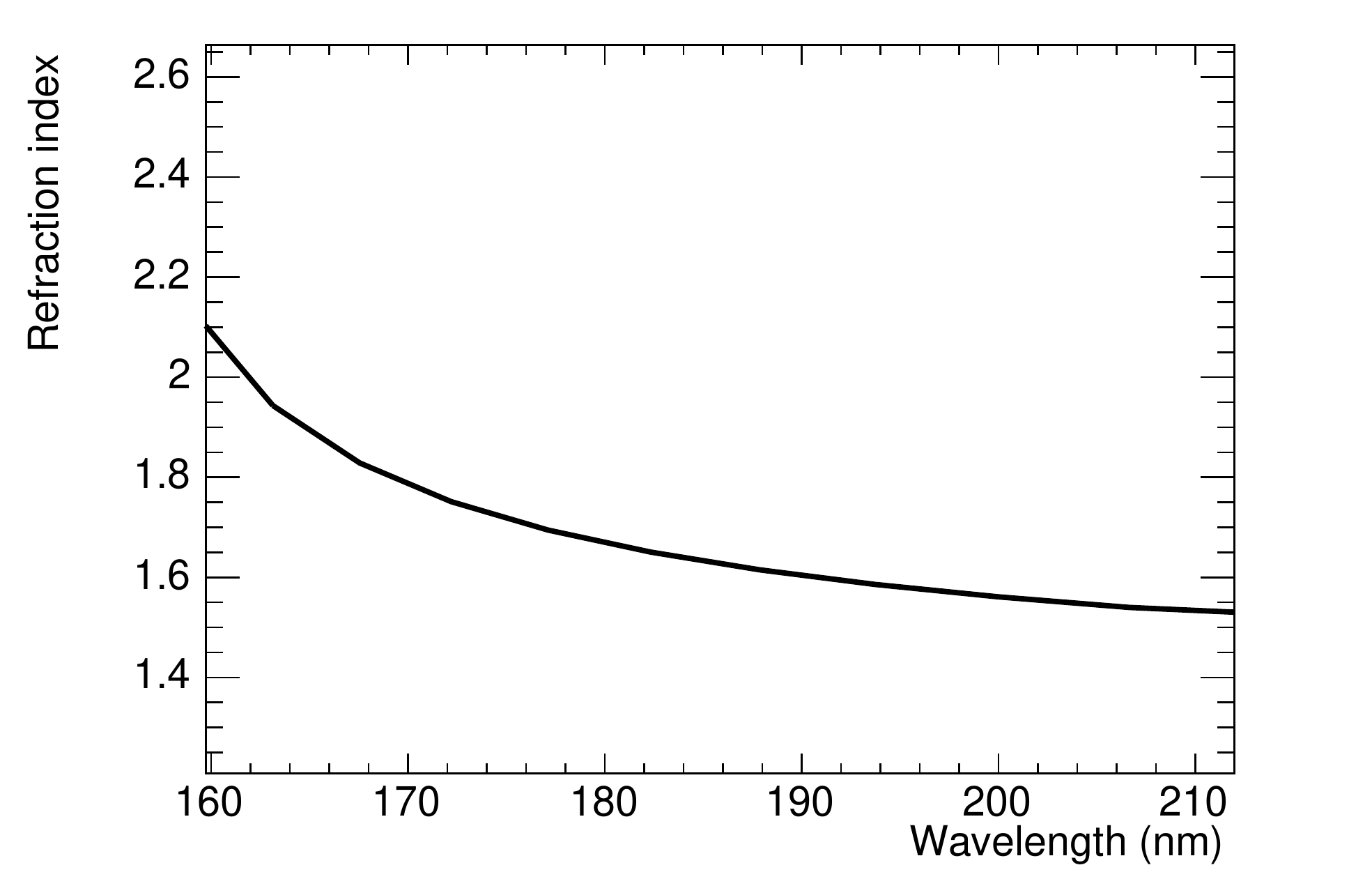}
	\caption{\label{fig.nlambda} LXe refraction index as a function of the wavelength of the optical photon, as results from the parametrization in reference \cite{Baldini:2004td}. For wavelengths in the emission range of TPB ($\sim$ 400--600 nm) the refraction index is practically flat, and has a value of 1.4.}
\end{figure}

Notice that, while the calculation of the term $\Delta d_g/c$~is immediate, finding the value
of $\Delta d_p/v_p$ requires the knowledge of the refraction index of the medium (which in turn is a function of the wavelength of the scintillation photons, see figure \ref{fig.nlambda}) and the interaction vertices of the gammas in the LXSCs, whose determination depends on the spatial resolution (and in particular of the DOI resolution) of the setup. Furthermore, the recorded time of the photoelectrons is affected by the time jitter of the sensor and the front-end electronics, as well as by the overall clock system. Other effect that affect the CRT is the additional decay constant (about 2.2 ns) introduced if the SiPMs of the LXSC2 are coated with TPB. 

In the remaining of this section we study systematically the various factors contributing to the CRT in LXe.

\subsection{Intrinsic CRT: effect of the DOI resolution}
We start our study by assuming that the refraction index of LXe is constant, with a value of 1.7 (which is the average of the values it assumes in the scintillation wavelength range). We also fix the refraction index of the SiPM window to the same value so that $n_1 = n_2$ and therefore all photons impinging the SiPM with an angle less than $90\,^{\circ}$~enter the sensor. These (unphysical) assumptions are made in order to separate the contributions to the CRT due to the scintillation lifetime and the DOI resolution from those related with the refraction index. 

\begin{figure}[!bhtp]
	\centering
	\includegraphics[scale=0.4]{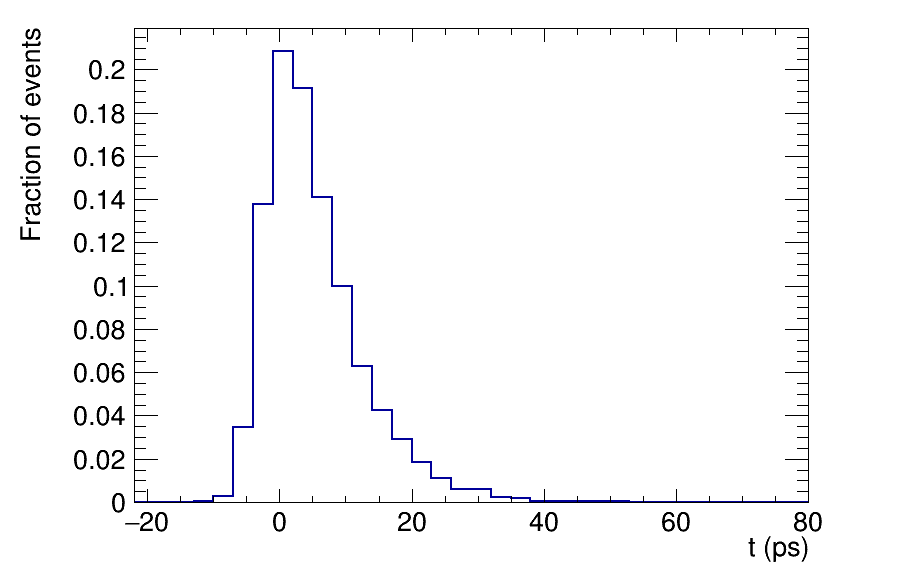}
	\caption{\label{fig.firstPE} Arrival time of the first recorded photon in LXe, after subtracting the time of flight of the incident gamma ($d_g/c$) and the time of flight of the scintillation photons ($n \cdot d_p/c$). The presence of negative bins is due to the the sum of two effects: on one hand, the time registered by sensors is binned with a size of 5 ps, while the time of flight of incident gammas is taken with the precision provided by the Monte Carlo (sub picosecond). On the other hand, the time of flight of the scintillating photon is determined by the position of the SiPM which detects it. Since SiPMs have an area of 3 $\times$ 3 $\textrm{mm}^2$, the error associated is of the order of $\sim$ 1.5 mm, which translates to $\sim$ 5 ps for velocities similar to the one of light.}\label{fig.FirstPE}
\end{figure} 

Figure \ref{fig.FirstPE} shows the arrival time of the first photoelectron, after subtracting the time of flight of the incident gamma ($d_g/c$) and the time of flight of the scintillation photons ($n \cdot d_p/c$). 
Notice that $d_p$~is the difference 
between the position of the interaction vertex and the position of the SiPM which records the photoelectron. 
The plot is computed introducing a resolution of 4.7 mm FWHM both in the transverse coordinates and
in the DOI (which is a conservative approach since a slightly better resolution is expected in the LXSC2). The jitter introduced by the DOI correction is
$
\delta \Delta t =\Delta z \times n \times 3.3\ \textrm{ps/mm}
$, which gives $\delta \Delta t = 26.3$~ps FWHM for a DOI resolution $\Delta z = 4.7$~mm FWHM. 

\begin{figure}[!bhtp]
	\centering
	\includegraphics[scale=0.4]{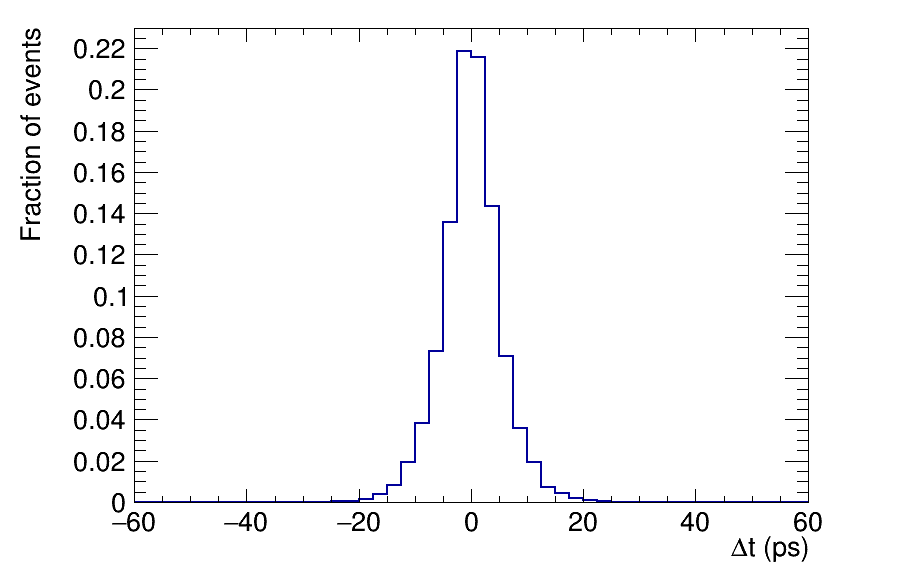}
	\caption{\label{fig.dtof} $\delta \Delta t$ in LXe. The
	histograms are obtained introducing a spatial resolution of 4.7 mm FWHM for the transverse coordinates and
	the DOI, fixing the wavelength of the scintillation and the refraction index and setting the PDE of the SiPMs to one. }
\end{figure}

\begin{figure}[!bhtp]
	\centering
	\includegraphics[scale=0.4]{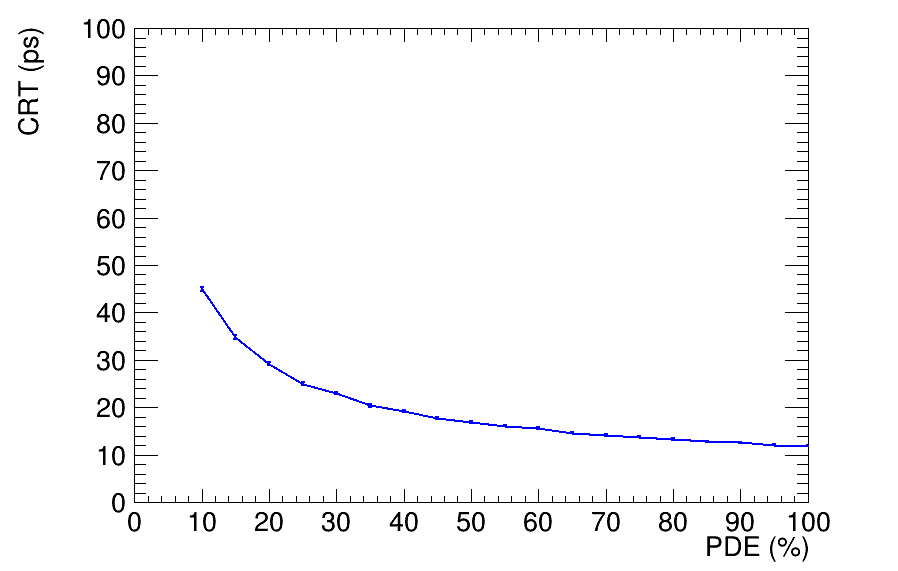}
	\caption{\label{fig.crt1} CRT as a function of the PDE computed using the first photoelectron. The calculation takes into account the scintillation yield and decay time as well as the DOI resolution but does not include
	the dependence of the refraction index in LXe with wavelength.}
\end{figure}

 Figure \ref{fig.dtof} shows $\delta \Delta t$ for the ideal case of PDE = 1. Figure \ref{fig.crt1} shows the CRT as a function of the
 SiPM PDE. The CRT for PDE = 1 is $\sim$26 ps (thus totally dominated by the uncertainty in DOI). This value can be taken as the absolute best limit of the CRT in LXe. For a PDE of 20\% (best value of VUV--sensitive SiPMs) the CRT in LXe is $\sim$ 30 ps.   

\subsection{Effect of the refraction index}
\begin{figure}[!bhtp]
	\centering
	\includegraphics[scale=0.6]{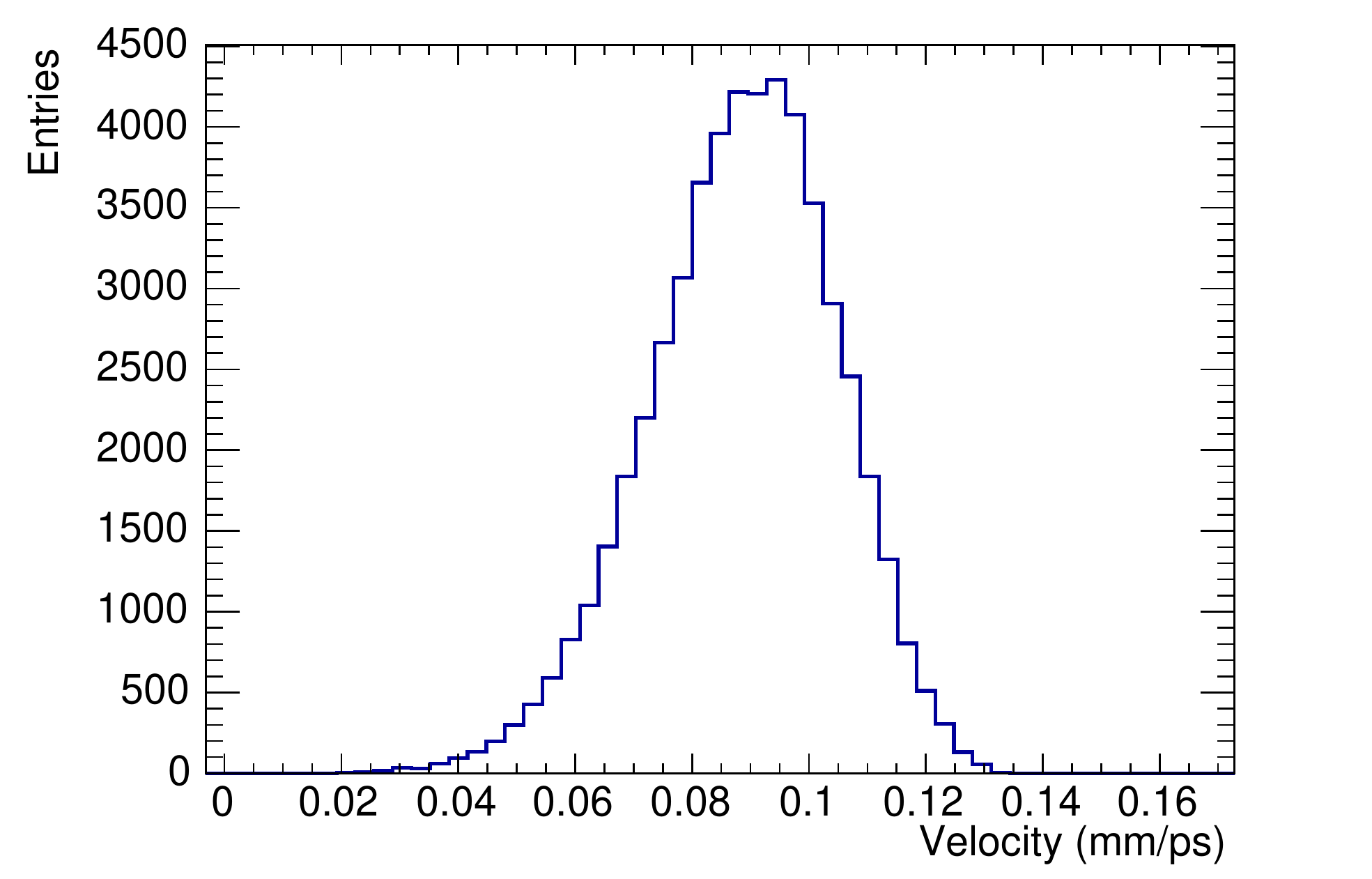}
	\caption{\label{fig.vLXe} Distribution of the velocity of scintillation photons propagating in LXe, given the refraction index shown in figure \ref{fig.nlambda} and the emission spectrum shown in figure \ref{fig.spectrumLXe} and considering the group velocity described in Eq.~\eqref{eq.velocity}.}
\end{figure}

The refraction index in LXe varies significantly with the scintillation photon wavelength, as shown in figure  \ref{fig.nlambda}.  The velocity of propagation of VUV photons in the Geant4 simulation is the group velocity $v_g$, rather than the phase velocity, and it depends on the refraction index, which, in turn, depends on the energy of the photon, according to the following relationship:
\begin{equation}
v_g = c\times \left(n(E)+\frac{\textrm{d}n}{\textrm{d}(\textrm{log}(E))}\right)^{-1}
\label{eq.velocity}
\end{equation}
This implies that the xenon VUV photons propagate with different velocities depending on their wavelength. The distribution of the velocity of  the optical photons emitted in LXe scintillation, simulated according to equation \eqref{eq.velocity}, is shown in figure \ref{fig.vLXe}. 
This introduces an extra smearing which contributes to the CRT, due to the spread of the average velocity. The measurement of the VUV photon velocity will be one of the subjects of study of the first PETALO prototype, to test the validity of this assumption.

We also introduce the refraction index of the SiPM entrance window, which we take as $n_1 = 1.54$ (a typical value, quoted, for example, for the SensL C-series 6 mm sensors \cite{pde}). The mismatch with the xenon refraction index results in a reduction of the efficiency due to Fresnel reflections. 

\begin{figure}[!bhtp]
	\centering
	\includegraphics[scale=0.40]{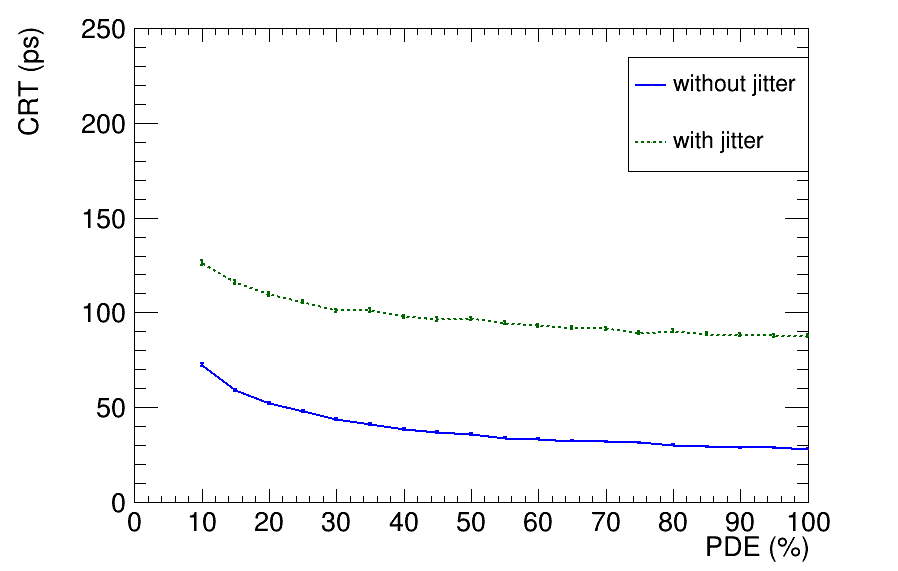}
	\caption{\label{fig.crt2} CRT as a function of the PDE computed using the first photoelectron. The calculation takes into account the scintillation yield and decay time as well as the DOI resolution and the dependence of the refraction index in LXe with wavelength. The effect of Fresnel reflections
	due to the mismatch between the LXe refraction index and that of the SiPM is also taken into
	account. The blue line does not include the effect of the SiPM and front-end eletronics jitter, while the green dotted line shows also this effect. }
\end{figure}

Figure \ref{fig.crt2} shows the CRT as a function of the
 SiPM PDE when the effect of the refraction index is taken into account. 
 The effect of the variable refraction index deteriorates considerably the intrinsic CRT in LXe, resulting in a value of $\sim$27 ps for PDE = 1. The CRT in LXe for a PDE of 20\% is $\sim$~50 ps.
  
\subsection{Effect of the SiPM  and front-end electronics jitter}
  Next we introduce the effect of the jitter of the SiPM and front-end electronics, smearing the arrival time of the photoelectrons by a gaussian distributions with $\sigma = \sigma_{\textrm{sipm}} +  \sigma_{\textrm{fee}}$ and setting 
 $\sigma_{\textrm{sipm}} = 80$~ps, $\sigma_{\textrm{fee}} = 30$~ps. These values are typical of modern SiPMs and front-end electronics \cite{FundamentalLimits}. 
 

 \begin{figure}[!bhtp]
	\centering
	\includegraphics[scale=0.40]{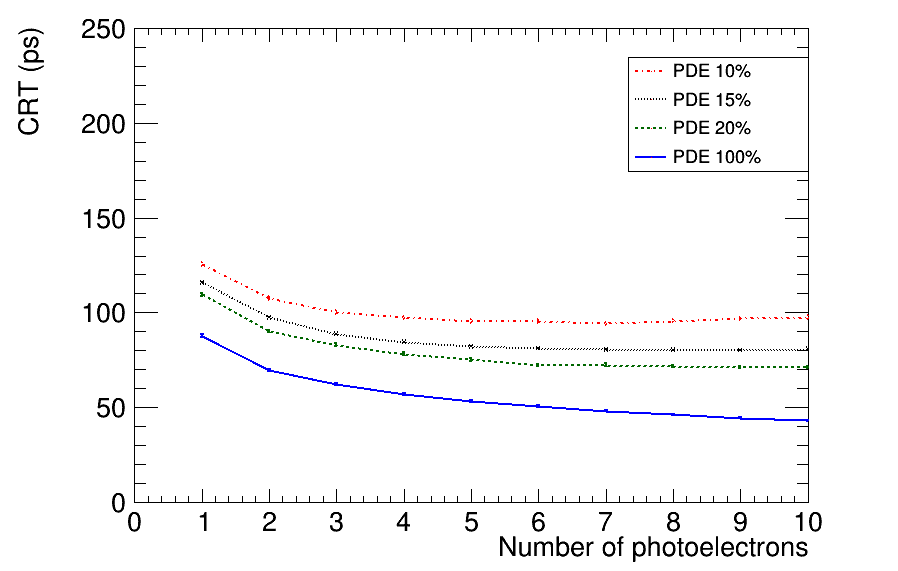}
	\caption{\label{fig.crt_avg_LXe} CRT as a function of the number of photoelectrons used to compute $\Delta t$. The CRT is shown for several values of the PDE.}
\end{figure}


Figure \ref{fig.crt2} shows the CRT as a function of the SiPM nominal PDE. The CRT for PDE = 1 is 90 ps, while for a PDE of 20\%  it is 110 ps. However, the CRT can improve computing $\Delta t$~as an average of the first few photoelectrons, rather than using only the first photoelectron. This is shown in figure \ref{fig.crt_avg_LXe}, which shows that for a PDE of 20\% it is possible to reach a
CRT of 70 ps.


\subsection{Effect of the TPB decay constant}

\begin{figure}[!bhtp]
	\centering
	\includegraphics[scale=0.4]{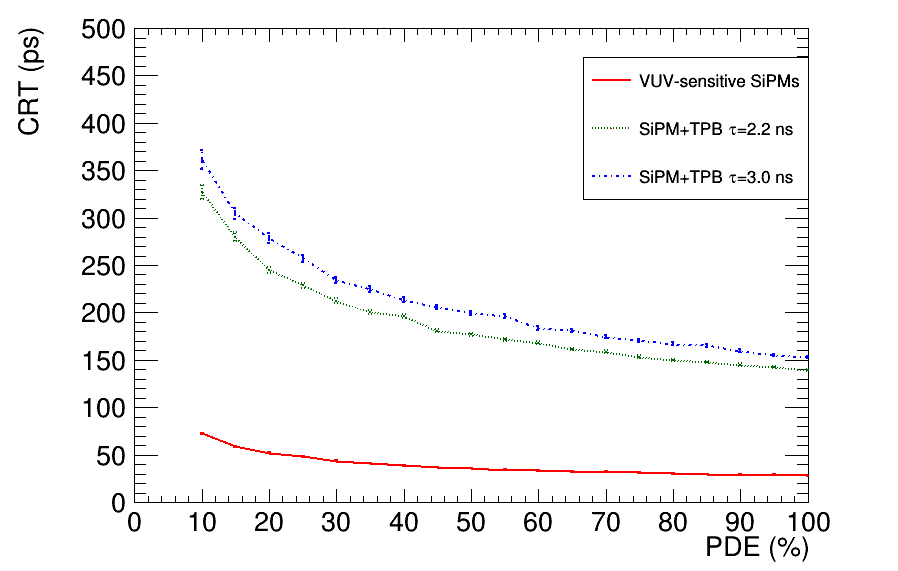}
	\caption{\label{fig.crtTPB} CRT as a function of the PDE using VUV-sensitive SiPMs and conventional SiPMs coated with TPB, assuming two decay times. The CRT is computed using the first photoelectron. The calculation takes into account the scintillation yield and decay time as well as the DOI resolution and the dependence of the refraction index in LXe with wavelength. The effect of the jitter of the SiPM and electronics is not included.}
\end{figure}

\begin{figure}[!bhtp]
	\centering
	\includegraphics[scale=0.4]{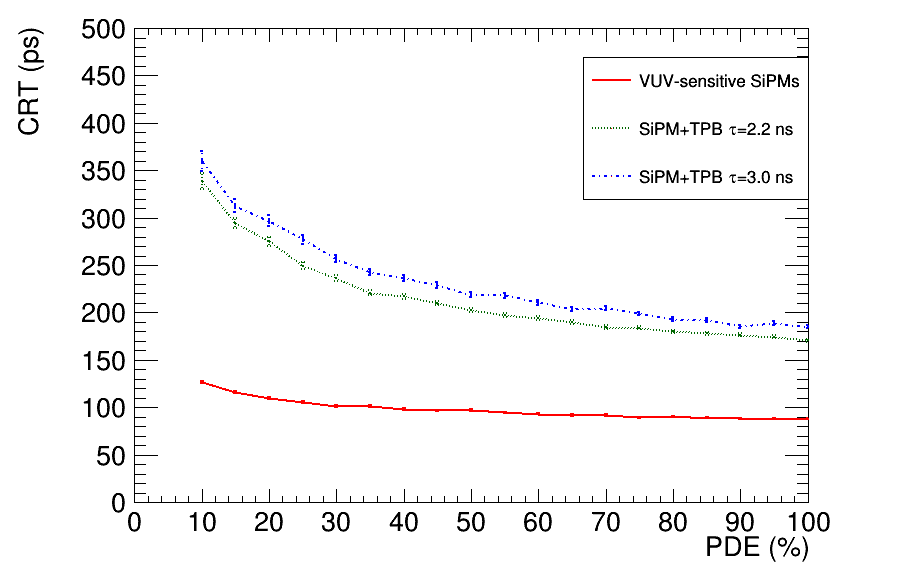}
	\caption{\label{fig.crtTPBjit} CRT as a function of the PDE using VUV-sensitive SiPMs and conventional SiPMs coated with TPB, assuming two decay times. The CRT is computed using the first photoelectron. The effect of the jitter of the SiPM and electronics is included. }
\end{figure}

\begin{figure}[!bhtp]
	\centering
	\includegraphics[scale=0.4]{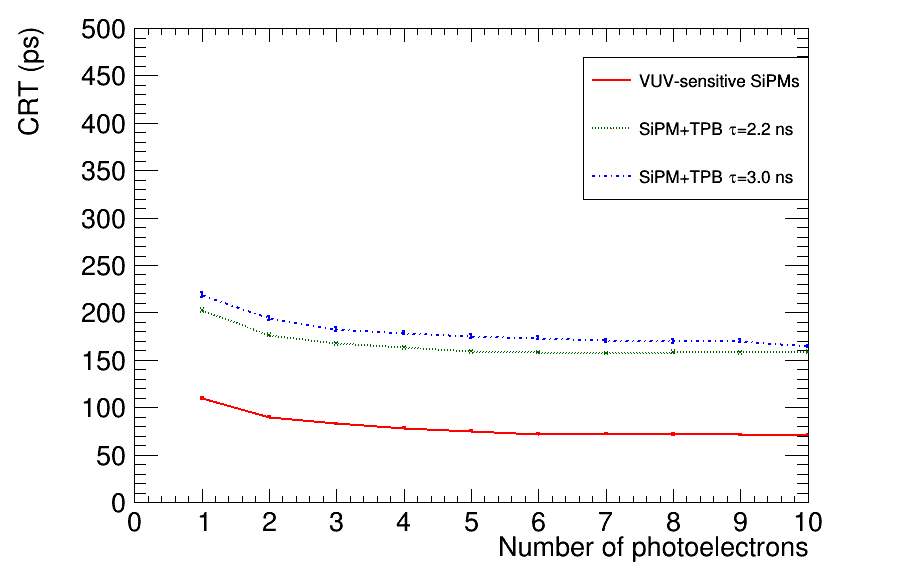}	
	\caption{\label{fig.crtTPBavg} CRT as a function of the number of photoelectrons used to compute $\Delta t$, for conventional SiPMs coated with TPB assuming a 50$\%$ PDE. The blue line is computed for a TPB decay time of 2.2 ns, while the green dotted line is for 3.0 ns. We also show the VUV-sensitive SiPM case for comparison.} 
\end{figure}

%
%

We now consider the case in which conventional SiPMs, with a PDE as large as 50\% in the blue region, are used in a PETALO scanner. The SiPMs would be coated with TPB to shift the VUV light to blue. The main reason to consider this scenario is cost. VUV sensitive SiPMs are currently much more expensive  than conventional SiPMs. It must be pointed out that such a cost difference can be quickly reduced with technological advances and massive production. Nevertheless, a PETALO scanner using TPB--coated SiPMs could be built with existing technology at a very competitive cost and is, therefore, interesting to study the effect of TPB on the CRT.

Figure \ref{fig.crtTPB} shows the CRT as a function of the
 SiPM nominal PDE for VUV sensitive SiPMs and TPB-coated SiPMs in LXe. 
 The calculation takes into account the scintillation yield and decay time as well as the DOI resolution and the dependence of the refraction index in LXe with wavelength, but not the jitter due to the SiPM and electronics. 
 Here we must compare the CRT achieved for VUV sensitive SiPMs with a PDE of 15--20\% ($\sim$ 50 ps) with the CRT achieved for blue sensitive SiPMs coated with TPB, taking as a reference a PDE of 50\% (this corresponds to the
 PDE for blue light,  however, the CRT is computed taking into account the effect of the TPB absorption and subsequent re-emission).  The CRT in this case varies from 180 ps (for a decay constant $\tau_{\rm{TPB}} = 2.2$~ns) to 
 200 ps (for a decay constant $\tau_{\rm{TPB}} = 3$~ns).  When the effect of the jitter of the SiPM and front-end electronics is added (figure \ref{fig.crtTPBjit}), the CRT (first photoelectron) for VUV-SiPMs is $110$~ps, while the CRT for TPB-coated SiPMs varies from  200 ps (for a decay constant $\tau_{\rm{TPB}} = 2.2$~ns) to  220 ps (for a decay constant $\tau_{\rm{TPB}} = 3$~ns). Averaging $\Delta t$ over the first 10 photoelectrons  the CRT is reduced to 160 ps for a decay constant $\tau_{\rm{TPB}} = 2.2$~ns  and 170 ps for a decay constant $\tau_{\rm{TPB}} = 3$~ns (figure \ref{fig.crtTPBavg}). Notice that averaging over several photoelectrons reduces the error in the CRT by a large factor in the case of SiPMs coated with TPB. 
 
 
 The results obtained with TPB are still very good: the best case scenario corresponds to a decay time of 2.2 ns and yields a CRT of 160 ps.
    
%
%

\section{Summary and outlook}\label{sec.conclu}

We have studied the CRT that can be obtained by a PETALO scanner based in the LXSC2 cell. The intrinsic resolution of the cell (corresponding to an ideal VUV sensitive sensor of PDE one and forcing a fixed refraction index) is $\sim$ 11 ps, totally dominated by the DOI uncertainty. Introducing a PDE of 20\%, corresponding to the best one currently achieved by VUV-sensitive SiPMs, still results in an excellent CRT of  30 ps. Including the effect of the variable refraction index and Fresnel reflections, a CRT of 50 ps is found for a PDE of 20\%. Adding the effect of the jitter of the SiPM and electronics spoils the CRT to some 110 ps for the same PDE. Averaging $\Delta t$~over the first few photoelectrons, one recovers a CRT of 70 ps. It follows that a PETALO scanner based in the LXSC2 cell equipped with VUV sensitive SiPMs of relatively large PDE (20\%) could reduce the CRT of a large system well below the 100 ps threshold, improving by a factor $\sim$ 4 the CRT of the best existing PET-TOF commercial system. 

On the other hand, we find that using blue sensitive SiPMs coated with TPB results in a CRT that could be as good as
160 ps (for $\tau_{\rm{TPB}} = 2.2$~ns). 
 In summary, it appears feasible to build a PETALO scanner with existing technology, combining the advantages of large sensitivity, good energy and spatial resolution, and excellent CRT.  

\acknowledgments
The authors acknowledge support from the following agencies and institutions: the European Research Council (ERC) under the Advanced Grant 339787-NEXT, the Ministerio de Econom'a y Competitividad and FEDER of Spain, the Severo Ochoa Program SEV-2014-0398 and GVA under grant PROMETEO/2016/120; we acknowledge enlightening discussions with J. Varela and C. Lerche.

\bibliography{biblio}

\end{document}